\documentclass[prb,twocolumn,showpacs]{revtex4}
\usepackage{amssymb}
\usepackage{graphicx}
\usepackage{amsfonts}
\usepackage{amsmath}

\begin{document}

\title{Parasitic pumping currents in an interacting quantum dot}

\author{F. Romeo and R. Citro}
\affiliation{$^{1}$Dipartimento di Fisica ''E. R. Caianiello'' and
Institute CNR-SPIN, Universit{\`a} degli Studi di Salerno, Via Ponte don Melillo,
I-84084 Fisciano (Sa), Italy}

\begin{abstract}
We analyze the charge and spin pumping in an interacting dot within the almost adiabatic limit.
By using a non-equilibrium Green's function technique within the time-dependent slave boson approximation, we analyze the pumped current
in terms of the dynamical constraints in the infinite-$U$ regime. The results show the presence of parasitic pumping currents due to the additional phases of the constraints. The behavior of the pumped current through the quantum dot is illustrated in the spin-insensitive and in the spin-sensitive case relevant for spintronics applications.
\end{abstract}

%73.23.-b electron transport in mesoscopic system
%72.25.Pn spin pump current-driven
%75.60.Jk magnetization reversal
%72.15.Qm scattering in electron transport metals


\maketitle

\section{Introduction}
The quantum pumping proposed by Thouless\cite{thouless} is a phase coherence effect able to pump dc current by the out-of-phase adiabatic modulation of (at least) two system parameters in absence of a voltage bias. The phase difference $\varphi$ of the external signals produces a charge current proportional to $\omega \sin(\varphi)$, $\omega$ being the adiabatic pumping frequency. After
the Thouless theory, Brouwer\cite{brouwer} formulated a scattering approach to the adiabatic
quantum pumping showing that the dc charge current produced by the pump is related to the parametric derivatives of the
scattering matrix of the device.

Following the Brouwer formulation several studies, both of experimental\cite{exp-pump} and theoretical nature, have been performed.
The theory of quantum pumping for noninteracting systems\cite{non-int-pump} has been developed in several directions.
In case of
interacting electrons the computation of the pumped charge becomes
rather involved and few works have addressed this issue for
different systems\cite{various_interacting} and in specific
regimes. As for the case of interacting quantum dots, the pumped
charge in a period was calculated by Aono\cite{aono_qd} by
exploiting the zero-temperature mapping of the Kondo problem  within non-equilibrium Green's function (NEGF) technique. A
very general formalism was developed in Ref.[\onlinecite{fazio_prl_05}]
based on the adiabatic expansion of the self-energy in the
average-time approximation, while a linear response scheme was employed in
Ref.[\onlinecite{sela_prl_06}]. Another interesting
study\cite{silva_07} was performed aiming at generalizing
Brouwer's formula for interacting systems to include inelastic
scattering events.
Alternatively to the NEGF approach, a powerful diagrammatic technique to treat the interacting quantum pumping is constituted by a generalized-master equation method\cite{konig_diagrammatic}.

Very recently the idea that the presence of an hidden classical variable inside a scattering region can strongly affect the dependence of the current pumped through the system has been proposed. This effect is produced by the dynamics of the classical variable which modifies the standard phase relation introducing a dynamical phase shift\cite{deformable-qd, memory-pumping} accompanied by \textit{parasitic} contributions to the pumping current.

In this paper we develop a time dependent mean field slave boson theory for an interacting quantum dot following a non-equilibrium formulation \textit{\`{a } la } Keldysh. In particular,
we derive a time-dependent mean field equation for the slave boson and the constraints
 which allows us to study the strongly interacting regime of the pumping beyond the adiabatic (i.e. zero frequency) limit. However, since the cost of the analysis is growing with the degree of non-adiabaticity, in this work we focus on the almost adiabatic limit (i.e. pumping frequency small but finite). In particular we focus on the effect of the dynamical phase of the slave boson on the pumped current, evidencing features not discussed before.
 Furthermore, the proposal of a quantum pump able to produce spin current without polarized electrodes is also discussed.

The organization of the paper is the following:
In Sec.\ref{sec:model}, we introduce the model Hamiltonian and derive the free-fermions-like effective Hamiltonian by means of the slave boson treatment.
In Sec.\ref{sec:negf} we derive the relevant Green's functions within the Keldysh approach and a formula for the pumping current is derived within the single photon approximation. In Sec.\ref{sec:constraints} we analyze the constraint equations and propose an approximate solution suitable within the almost adiabatic limit.
In Sec.\ref{sec:results} we discuss the results obtained both in the case of a spin independent quantum pump and in the case of a spin sensitive device able to generate spin current. The conclusions are given in Sec.\ref{sec:conclusions}.

\section{ The model and formalism}
\label{sec:model}
\begin{figure}
%\vspace{3.0cm}
\centering
\includegraphics[scale=0.35]{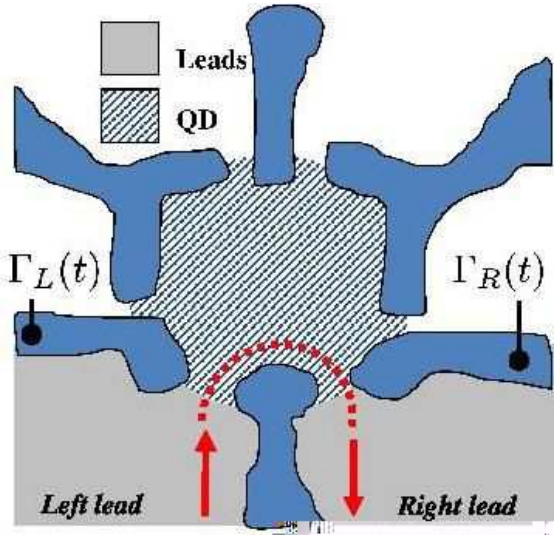}\\
\caption{The system described in the main text: A Quantum dot (dashed region) coupled to external leads via tunnel barriers whose transparency can be modulated in time via top gates. } \label{fig:fig1}
\end{figure}

We consider a quantum dot (QD)-based pump coupled to noninteracting leads. A particle flux is generated through the QD by means of an almost-adiabatic modulation of two out-of-phase gate voltages controlling the transparency of the tunneling barriers.\\
 The Hamiltonian of the quantum pump depicted in Fig.\ref{fig:fig1} can be written as follows:
\begin{eqnarray}
H&=&\sum_{k\sigma\alpha}\varepsilon_k c_{k\sigma\alpha}^{\dag}c_{k\sigma\alpha}+\sum_{\sigma}\epsilon \ d_{\sigma}^{\dag} d_{\sigma}+U n_{\uparrow}n_{\downarrow}+\\\nonumber
&+&\sum_{k\sigma\alpha}(V_{k\alpha\sigma}(t)c_{k\sigma\alpha}^{\dag}d_{\sigma}+h.c.),
\end{eqnarray}
where the operator $c^{\dag}_{k\sigma\alpha}$ creates an electron of momentum $k$ and spin $\sigma$ in the lead $\alpha$,
while $d^{\dag}$ is the creation operator of an electron state on the interacting quantum dot, $\epsilon$ being the energy of a single occupied electron state. The electron-electron (e-e) interaction energy $U$, described by the third term, has to be considered when the dot screening length $\lambda_s$ is bigger than the typical size of the quantum dot. The last term describes the tunneling between the quantum dot and the leads. In this work we are interested in describing the strongly interacting limit ($U \rightarrow \infty$)\cite{nota1}. In the intermediate $U$-regime the Hilbert space of the isolated QD has electron states
$\{|0\rangle, |\sigma\rangle, |2\rangle\}$ (empty, single and doubly occupied) which are created by application of the creation operator to the empty state $|0\rangle$, i.e. $|\sigma\rangle=d_{\sigma}^{\dag}|0\rangle$, $|2\rangle=|\uparrow\downarrow\rangle=d_{\uparrow}^{\dag}d_{\downarrow}^{\dag}|0\rangle$ (notice that the order is arbitrary but have to be fixed since $|\uparrow\downarrow\rangle=-|\downarrow\uparrow\rangle$). In the previous basis the relevant operators can be written as\cite{nota1b}:
\begin{eqnarray}
d_{\sigma'}^{\dag}&=&|\sigma'\rangle\langle 0|+\sigma'|2\rangle\langle\bar{\sigma}'|\\\nonumber
d_{\sigma}&=& |0\rangle\langle \sigma|+\sigma|\bar{\sigma}\rangle\langle2|\\\nonumber
n_{\sigma}&=& |\sigma \rangle\langle \sigma|+|2 \rangle\langle 2|\\\nonumber
n_{\uparrow}n_{\downarrow}&=&|2 \rangle \langle 2|,
\end{eqnarray}
and the canonical anti-commutation relations
$\{d_{\mu},d_{\nu}^{\dag}\}=\delta_{\mu\nu}$, $\{d_{\mu},d_{\nu}\}=0$, $\{d_{\mu}^{\dag},d_{\nu}^{\dag}\}=0$ can be obtained.
In case of strong  interaction ($U \rightarrow \infty$) the occupation of the $|2\rangle\langle2|$ subspace becomes strongly suppressed and thus the commutation relations\cite{leguillou-commutation} become $\{d_{\mu},d_{\nu}^{\dag}\}=\delta_{\mu\nu}-\mu\nu d^{\dag}_{\bar{\mu}}d_{\bar{\mu}}$,  where $\mu,\nu \in \{\uparrow,\downarrow\}\equiv\{+1,-1\}$ and the effective completeness relation is $1=|0\rangle\langle0|+\sum_{\sigma}|\sigma\rangle\langle\sigma|$. Thus the electron operators, with the exclusion of the doubly occupied state, take the form $d_{\sigma}=|0\rangle\langle\sigma|$, $d_{\sigma}^{\dag}=|\sigma \rangle\langle0|$. By introducing the reference state $|ref\rangle$ and the operators $b$ and $f_{\sigma}$ we define $|0\rangle=b^{\dag}|ref\rangle$ and $|\sigma\rangle=f_{\sigma}^{\dag}|ref\rangle$ and thus the creation/annihilation operators can be written in terms  of bosonic (i.e. $b$) and quasi-fermionic (i.e. $f_{\sigma}$) operators\cite{coleman, finnemore} as: $d_{\sigma}\rightarrow b^{\dag}f_{\sigma}$, $d_{\sigma}^{\dag}\rightarrow f_{\sigma}^{\dag}b$.
Within the slave boson representation the Hamiltonian is:
\begin{equation}
H_{SB}=H(d_{\sigma}\rightarrow b^{\dag}f_{\sigma};d_{\sigma}^{\dag}\rightarrow f_{\sigma}^{\dag}b)+\lambda (b^{\dag}b+\sum_{\sigma}f_{\sigma}^{\dag}f_{\sigma}-1).
\end{equation}
where the single occupation constraint is included by the Lagrange multiplier $\lambda$ which is fixed by the equation:
\begin{equation}
\partial_{\lambda}\langle H_{SB}\rangle=0 \rightarrow \sum_{\sigma}\langle f^{\dag}_{\sigma}f_{\sigma}\rangle+\langle b^{\dag}b\rangle-1=0,
\end{equation}
while the slave boson operator $b$ evolves in time according to the equation of the motion $i\hbar \partial_t b=[b,H_{SB}]$, thus\cite{wu_tdsb}:
\begin{equation}
i\hbar\partial_t b=\lambda b+\sum_{k\sigma\alpha}V_{k\alpha\sigma}(t)c^{\dag}_{k\sigma\alpha}f_{\sigma}.
\end{equation}
We treat the slave boson operator $b$ within the mean field approximation ($\langle b \rangle=\mathcal{B}$, $\langle b^{\dag}\rangle= \mathcal{B}^{\ast}$) and the original problem containing strong correlations is replaced by a constrained free-fermions-like theory whose dynamics is completely described by the following Hamiltonian:
\begin{eqnarray}
H_{SBMF} &=& H_{leads}+\sum_{\sigma}(\epsilon+\lambda(t))f^{\dag}_{\sigma}f_{\sigma}+\\\nonumber
&+&\sum_{k\sigma\alpha}(V_{k\alpha\sigma}(t)\mathcal{B}^{\ast}(t) c^{\dag}_{k \sigma\alpha}f_{\sigma}+h.c.)+\\\nonumber
&+&\lambda(t)(|\mathcal{B}(t)|^2-1),
\end{eqnarray}
with the dynamical constraints:
\begin{eqnarray}
\label{eq:b-dynamics}
i\hbar\partial_t \mathcal{B}(t)=\lambda(t) \mathcal{B}(t)+\sum_{k\sigma\alpha}V_{k\alpha\sigma}(t) \langle c^{\dag}_{k\sigma\alpha}f_{\sigma}\rangle\\
\label{eq:b-dynamics2}
\sum_{\sigma}\langle f^{\dag}_{\sigma}f_{\sigma}\rangle+|\mathcal{B}(t)|^2-1=0.
\end{eqnarray}
As shown in (6) the interaction produces a renormalization of the QD energy level ($\epsilon \rightarrow \epsilon + \lambda(t)$) and a redefinition of the tunneling amplitudes ($V_{k\alpha\sigma}(t) \rightarrow V_{k\alpha\sigma}(t)\mathcal{B}^{\ast}(t)$). This aspect is very important in the context of the quantum pumping.
In fact, since the time evolution of the slave boson field is governed by the equation (\ref{eq:b-dynamics}), a dynamical phase shift with respect to the phase of the external driving signals is expected. Within the adiabatic case, all the phases are rigidly related to the external driving signals, while at finite frequency this relation is lost and interesting contribution of hidden phases can emerge.

\section{NEGF method}
\label{sec:negf}

 In this section we employ the non-equilibrium Green's functions (NEGF) formalism\cite{meir-wingreen_NEGF-method} to derive the current flowing through the system when subject to an almost adiabatic quantum pumping (small but finite pumping frequency $\omega$) in which we modulate the tunneling rates $\Gamma_{\sigma}^{\alpha}(t)=\Gamma_{\sigma,0}^{\alpha}+\Gamma_{\sigma,\omega}^{\alpha}\sin(\omega t+\varphi_{\alpha})$ defined by $\Gamma_{\sigma}^{\alpha}(t)=2\pi\rho_{\alpha}|V_{k_F\alpha\sigma}(t)|^2$, where $\rho_{\alpha}$ is the density of states at the Fermi level $k_F$.\\
 Due to the free-fermions like form of the Hamiltonian the retarded GF of the quantum dot (QD) uncoupled from the external leads is ($\hbar=1$):
\begin{equation}
g_{sp}^{r}(t,t')=-i\delta_{sp}\theta(t-t')\exp\{-i\int_{t'}^{t}dt_1(\epsilon+\lambda(t_1))\},
\end{equation}
being $s,p\in \{\uparrow,\downarrow \}$ the spin index. When the QD is coupled to the leads, we must take into account the transition rate of an electron through the system. The quantity encoding this information is the retarded self-energy which can be written within the wide band limit (WBL) as:
\begin{equation}
\Sigma_{sp}^{r}(t_1,t_2)=-i\delta_{sp}\delta(t_1-t_2)\sum_{\alpha=l,r}\frac{\Gamma_s^{\alpha}(t_1)}{2}|\mathcal{B}(t_1)|^2,
\end{equation}
 and thus the retarded Green's function of the QD coupled to the leads takes the form:
 \begin{equation}
 G_{sp}^{r}(t,t')=g_{sp}^{r}(t,t')\exp\{ -\int_{t'}^{t}dt_1\sum_{\alpha=l,r}\frac{\Gamma_s^{\alpha}(t_1)}{2}|\mathcal{B}(t_1)|^2\},
 \end{equation}
 which depends on the renormalized quantities $|\mathcal{B}(t)|^2$ and $\lambda(t)$. The corresponding advanced quantities, i.e. $G^a$ and $\Sigma^a$, are computed directly by using the general relation $\Xi_{sp}^a(t_1,t_2)=\Xi_{ps}^r(t_2,t_1)^{\ast}$, $\Xi$ being $G$ or $\Sigma$. Finally, the Langreth rules\cite{langreth-theorem} can be employed to compute the \textit{lesser} self-energy as
 \begin{equation}
\Sigma_{sp}^{<}(t_1,t_2)=i\delta_{sp}f(t_1-t_2)\sum_{\alpha=l,r}\Gamma_s^{\alpha}(t_1)|\mathcal{B}(t_1)|^2.
\end{equation}
It is proportional to the Fourier transform $f(t_1-t_2)$ of the Fermi function $f(E)$ (notice that the chemical potential is the same for both the leads).

In order to calculate the current flowing through the QD we need to calculate the \textit{lesser} GF $G_{sp}^{<}(t_1,t_2)$ of the QD exploiting the Keldysh equation\cite{haug_book} and by using $G^{r/a}$ and $\Sigma^{<}$.
 Together with the single particle Green's function we need yet to solve the constraint equation.
Explicitly we can rewrite Eq.(\ref{eq:b-dynamics}) and (\ref{eq:b-dynamics2}) as:
 \begin{eqnarray}
 \label{eq:constraints-GF1}
 i\partial_t\mathcal{B}(t)&=&\Bigl[\lambda(t)+\sum_{\sigma}G^{<}_{\sigma\sigma}(t,t)\frac{\Gamma_{\sigma}(t)}{2}\Bigl]\mathcal{B}(t)+\\\nonumber
 &+&\sum_{\sigma}\int dt_1 G^{r}_{\sigma\sigma}(t,t_1)\mathcal{B}(t_1)\Gamma_{\sigma}(t_1)f(t_1-t)\\
\label{eq:constraints-GF2}
|\mathcal{B}(t)|^2 &=& 1+i\sum_{\sigma}G^{<}_{\sigma\sigma}(t,t)
 \end{eqnarray}
%Previous equations jointly with the $H_{SBMF}$ completely describe the physics of the interacting quantum dot in the infinite-$U$ limit.
Let us note that the interaction effects are taken into account by the constraints. This aspect becomes evident considering the limiting case in which $\Gamma_{\sigma}=\Gamma$ (spin insensitive tunneling). Solving (\ref{eq:constraints-GF2}) for $\sum_{\sigma}G_{\sigma\sigma}^{<}$ and substituting into (\ref{eq:constraints-GF1}), we obtain
\begin{equation}
i\partial_t \mathcal{B}(t)=[\lambda(t)+i\frac{\Gamma(t)}{2}]\mathcal{B}(t)-i\frac{\Gamma(t)}{2}|\mathcal{B}(t)|^2\mathcal{B}(t)+...,
\end{equation}
in which the nonlinear term $|\mathcal{B}|^2\mathcal{B}$ accounting for the interacting nature of the problem appears.
 The obtained equations are similar the one derived by Langreth \textit{et al.} in Ref.[\onlinecite{langreth_nordlander}] even though in our time-dependent theory we expect an easier manipulation.

\subsection{Non-equilibrium currents}
Within the non-equilibrium theory the spin resolved current pumped in the lead $\alpha$ can be written in terms of the QD Green's function in the following form:
\begin{eqnarray}
\vspace{0.5cm}
I_{\alpha}^{\sigma}(t)&=&\frac{2e}{\hbar}\sum_{\sigma'}Re\Bigl\{ \int \frac{dE_1 dE_2 dE_3}{(2\pi)^3}\\\nonumber
&\times &\exp[i(E_3-E_1)t][G^{r}_{\sigma\sigma'}(E_1,E_2)\Sigma^{<}_{\alpha,\sigma'\sigma}(E_2,E_3)+\\\nonumber
&+& G^{<}_{\sigma\sigma'}(E_1,E_2)\Sigma^{a}_{\alpha,\sigma'\sigma}(E_2,E_3)]\Bigl\},
\end{eqnarray}
where the two-times Fourier transform\cite{nota2} of the GF have been introduced. In this work we are interested in deriving the pumping current generated by an almost-adiabatic perturbation and thus we expand all the relevant time dependent quantities by considering only one-photon contributions as follows:
\begin{eqnarray}
\lambda(t)&=&\sum_{n=0,\pm 1} \lambda_{n}\exp\{-i n \omega t\}\\\nonumber
\Gamma_{\sigma}^{\alpha}(t)&=&\sum_{n=0,\pm 1} \Gamma_{\sigma;n}^{\alpha}\exp\{-i n\omega t\}\\\nonumber
\mathcal{B}(t)&=&\sum_{n=0,\pm 1}\mathcal{B}_{n}\exp\{-i n \omega t\}\\\nonumber
\Xi^{\xi}_{sp}(E_1,E_2)&=&\delta_{sp}\sum_{\eta=0,\pm1}\Xi^{\xi}_{s;\eta}(E_1)\delta(E_1-E_2+\eta \omega),
\end{eqnarray}
where $\Xi$ is $G$ or $\Sigma$, while $\xi \in \{r, a, < \}$. Within the one-photon approximation, the spin resolved dc current $\bar{I}^{\sigma}_{\alpha}$ pumped in the lead $\alpha$ takes the form $\bar{I}^{\sigma}_{\alpha}=(2e/\hbar)\sum_{\eta=\pm 1}Re[\mathcal{A}_{\sigma;\eta}^{\alpha}]$ where we defined the following quantity:
\begin{eqnarray}
\mathcal{A}_{\sigma;\eta}^{\alpha}&=&\int\frac{dE_1}{(2\pi)^3}[G^{r}_{\sigma;\eta}(E_1)\Sigma^{<}_{\alpha\sigma;-\eta}(E_1+\eta\omega)+\\\nonumber
&+&G^{<}_{\sigma;\eta}(E_1)\Sigma^{a}_{\alpha\sigma;-\eta}(E_1+\eta\omega)],
\end{eqnarray}
which contains information about the pumping cycle and the absortion/emission processes of one photon.
Within the single photon approximation the behavior of the pumped current with respect to the QD energy level is determined by the two integrals:
\begin{eqnarray}
\mathcal{R}_1^{\sigma\eta}=\int_{\Lambda}\frac{dE}{2\pi}f(E)\mathcal{D}^{\sigma}_{\eta}(E)\mathcal{D}^{\sigma}_0(E)\\\nonumber
\mathcal{R}_2^{\sigma\eta}=\int_{\Lambda}\frac{dE}{2\pi}f(E)\mathcal{D}^{\sigma}_{-\eta}(E)\mathcal{D}^{\sigma}_0(E)^{\ast},
\end{eqnarray}
where
\begin{equation}
\mathcal{D}^{\sigma}_{\eta}(E)=[E+\eta\omega -(\epsilon+\lambda_0)+i(\gamma_{\sigma}/2)]^{-1},
\end{equation}
while $\lambda_0$ represents the static part of the Lagrange multiplier and $\gamma_{\sigma}$ is a renormalized linewidth depending on the slave boson field:
\begin{eqnarray}
\label{eq}
\gamma_{\sigma}=\sum_{\alpha}\Bigl[|\mathcal{B}_0|^2\Gamma_{\sigma;0}^{\alpha} &+& 2Re\{|\mathcal{B}_0|^2\Gamma_{\sigma;1}^{\alpha}+\\\nonumber
&+&\Gamma_{\sigma;0}^{\alpha}(\mathcal{B}^{\ast}_0\mathcal{B}_1+\mathcal{B}_0\mathcal{B}^{\ast}_{-1})\}\Bigl].
\end{eqnarray}
%===================================================================fig1
%=======================================================================
 As shown in (\ref{eq}) the finite frequency effects modifies the mean lifetime of the electron on the QD due to the interference effects originated by the slave boson field $\mathcal{B}_{\pm 1}$. Furthermore, since $\Gamma_{\sigma;\pm 1}^{\alpha}$ picks up a phase due to the external driving signals $\Gamma_{\sigma;\pm 1}^{\alpha}=\pm i\frac{\Gamma_{\sigma,\omega}^{\alpha}}{2}\exp\{\mp i \varphi_{\alpha}\}$, the GF of the problem depends separately on $\varphi_{\alpha}$ and not via the phase difference $\varphi=\varphi_R-\varphi_L$. The last observation implies that the quantum state of the dot is affected by the precise external phases which couple the system and not only by the phase difference $\varphi$ applied to the nanostructure. Moreover, since the slave boson field is a complex quantity,  $\mathcal{B}_{\pm1,0}=|\mathcal{B}_{\pm1,0}|\exp\{i \phi_b^{\pm,0}\}$, interference terms, i.e. of the form $\cos(\phi_b^{\pm}-\phi_b^{0})$ appears in $\mathcal{D}_{\eta}^{\sigma}(E)$. Also additional phases associated to the constraint evolution appear. Thus beyond the almost adiabatic limit, the existence of a great number of dynamical phases (and the corresponding interference terms) provides a decoherence source for the electron transport.

\section{Constraints and their approximate solution}
\label{sec:constraints}

\subsection{Constraints within the single-photon approximation}
The expression of the current pumped through the system can be calculated once the GF is determined. Since it depends parametrically on $\{\lambda_0,\lambda_{\pm1};\mathcal{B}_0,\mathcal{B}_{\pm1}\}$ one has first to solve the constraint equations. By exploiting the Fourier expansions of Eqs.(\ref{eq:constraints-GF1})-(\ref{eq:constraints-GF2}) within the one-photon approximation we get the following set of constraints:
\begin{eqnarray}
\label{eq:constr-GF-fourier1}
&&\lambda_0 \mathcal{B}_0 = -\Phi^{(0;0)}(E=0)\\\nonumber
&&(\lambda_0-\omega)\mathcal{B}_1+\lambda_1\mathcal{B}_0 = -\sum_{\mu=0,1}\Phi^{(1-\mu;-\mu)}(E=\omega)\\\nonumber
&&(\lambda_0+\omega)\mathcal{B}_{-1}+\lambda_{-1}\mathcal{B}_0 = -\sum_{\mu=0,1}\Phi^{(-\mu;1-\mu)}(E=-\omega),
\end{eqnarray}
and
\begin{eqnarray}
\label{eq:constr-GF-fourier2}
&&|\mathcal{B}_0|^2=1-\sum_{\sigma}\int \frac{dE}{i(2\pi)^2}G^{<}_{\sigma;0}(E)\\\nonumber
&&\mathcal{B}^{\ast}_0\mathcal{B}_1+\mathcal{B}_0\mathcal{B}^{\ast}_{-1}=-\sum_{\sigma}\int \frac{dE}{i(2\pi)^2}G^{<}_{\sigma;-1}(E)\\\nonumber
&&\mathcal{B}^{\ast}_0\mathcal{B}_{-1}+\mathcal{B}_0\mathcal{B}^{\ast}_{1}=-\sum_{\sigma}\int \frac{dE}{i(2\pi)^2}G^{<}_{\sigma;1}(E),
\end{eqnarray}
where the auxiliary functions $\Phi^{(m;n)}(E)$ involving the GFs of the QD are reported in the appendix \ref{app:fi-function}.
 In principle the finite frequency problem gives us an infinite set of coupled constraints, while within the one photon approximation we obtain a reduced number of equations. However, from the numerical point of view, the solution of Eqs.(\ref{eq:constr-GF-fourier1})-(\ref{eq:constr-GF-fourier2}) may be in principle a non-trivial task since there can be numerical instabilities in the determination of the solution.
When an almost adiabatic regime is considered (retaining a finite frequency $\omega$) one considers the static constraints as perturbed by the remaining ones which on their turn can be treated by using some suitable approximations to be checked at the end of the computation.

\subsection{Approximate solution of the constraints}
In this subsection an approximate solution of the constraints shown in Eqs.(\ref{eq:constr-GF-fourier1})-(\ref{eq:constr-GF-fourier2}) is provided within the almost adiabatic case. From the analysis of the parametric expression of the current pumped through the system one can observe that the relevant quantities to be computed are: $\lambda_0$, $\lambda_1$, $|\mathcal{B}_0|^2$ and $\mathcal{B}^{\ast}_0\mathcal{B}_1+\mathcal{B}_0\mathcal{B}^{\ast}_{-1}$. Since we are modulating the tunneling amplitudes via top gates the contribution to the current coming from the modulation of $\lambda(t)$ can be considered as a \textit{parasitic pumping effect} whose intensity is very weak when $\lambda_1 \rightarrow 0$. Thus we need only to produce an approximate expression of $\lambda_{\pm 1}$ and $\mathcal{B}^{\ast}_0\mathcal{B}_1+\mathcal{B}_0\mathcal{B}^{\ast}_{-1}$ to the first order in $\Gamma_{\sigma;1}$. The constraints which need strictly to be valid are:
\begin{eqnarray}
\label{eq:constraints-GF-static}
\lambda_0=-\sum_{\sigma}\Gamma_{\sigma;0}\int\frac{d E}{2\pi}f(E)Re\{\mathcal{D}_0^{\sigma}(E)\}\\\nonumber
|\mathcal{B}_0|^2=1-\sum_{\sigma}|\mathcal{B}_0|^2\Gamma_{\sigma;0}\int\frac{d E}{2\pi}f(E)|\mathcal{D}_0^{\sigma}(E)|^2.
\end{eqnarray}
The previous equations come from the first equation of each set of constraints given in Eqs.(\ref{eq:constr-GF-fourier1})-(\ref{eq:constr-GF-fourier2}). They have the same structure as in the adiabatic case\cite{fazio_prl_05} (not presented here), even though, differently from that case, an hidden dependence on  $\mathcal{B}_{\pm 1}$ is present in $\mathcal{D}_0^{\sigma}(E)$ and thus higher order constraints have to be considered in order to solve the problem. In particular, by considering the second equation in (\ref{eq:constr-GF-fourier2}) it is possible to formally solve it in the following form:
\begin{eqnarray}
\label{eq:constr-bob1}
&&\mathcal{B}^{\ast}_0\mathcal{B}_1+\mathcal{B}_0\mathcal{B}^{\ast}_{-1}=\\\nonumber
&&=\frac{-\sum_{\sigma}|\mathcal{B}_0|^2\Gamma_{\sigma;1}\int\frac{dE}{2\pi}\mathcal{D}_0^{\sigma}(E)\mathcal{D}_{-1}^{\sigma \ast}(E)f^{(-)}(E)}{1+\sum_{\sigma}\Gamma_{\sigma;0}\int\frac{dE}{2\pi}\mathcal{D}_0^{\sigma}(E)\mathcal{D}_{-1}^{\sigma \ast}(E)f^{(-)}(E)},
\end{eqnarray}
where the right hand side of the previous equation depends on $\lambda_0$, $|\mathcal{B}_0|^2$ and $\mathcal{B}_{\pm1}$, while we defined $f^{(\pm)}(E)=f(E\pm\omega)$. Expanding the r.h.s. of Eq.(\ref{eq:constr-bob1}) up to the first order in $\Gamma_{\sigma;1}$ and taking the limit $\omega \rightarrow 0$, the dependence on $\mathcal{B}_{\pm1}$ disappears and thus Eqs.(\ref{eq:constraints-GF-static})-(\ref{eq:constr-bob1}) can be solved independently from the other equations. Following a similar procedure, considering the limit $\Gamma_{\sigma;0}/\lambda_0 \ll1$ we obtain:
\begin{equation}
\lambda_1 \approx \frac{\lambda_0}{2}\sum_{\sigma}|\mathcal{B}_0|^2\Gamma_{\sigma;1}\int\frac{dE}{2\pi}|\mathcal{D}_{0}^{\sigma}(E)|^2f(E)\Bigl|_{\mathcal{B}_{\pm1}\rightarrow0}.
\end{equation}
%It is worth to mention here that within the above approximations the phase of the quantities $\lambda_1$ and $\mathcal{B}^{\ast}_0\mathcal{B}_1+\mathcal{B}_0\mathcal{B}^{\ast}_{-1}$ is locked on the phase of the driving amplitude $\Gamma_{\sigma;1}$, while in general an additional phase cames from the integrals due to higher order effects.

\section{Results}
\label{sec:results}
\subsection{Spin-insensitive quantum pumping}
\begin{figure}
%\vspace{3.0cm}
\centering
\includegraphics[scale=0.75]{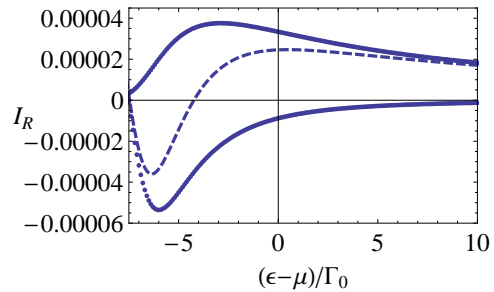}\\
\caption{Total Charge current $I_R$ (dashed line) as a function of $(\epsilon-\mu)/\Gamma_0$ computed setting the remaining parameters as follows:
$\Gamma_{\sigma;0}^{\alpha=L/R}=1.5$, $\Gamma_{\sigma;\omega}^{\alpha=L/R}=0.05$, $\omega=0.05$, $\varphi_L=0$, $\varphi_R=\pi/4$. The lower curve represent the current induced by the modulation of $\lambda$, while the upper curve depends only on the modulation of the tunneling amplitudes.} \label{fig:fig2}
\end{figure}

\begin{figure}
%\vspace{3.0cm}
\centering
\includegraphics[scale=0.7]{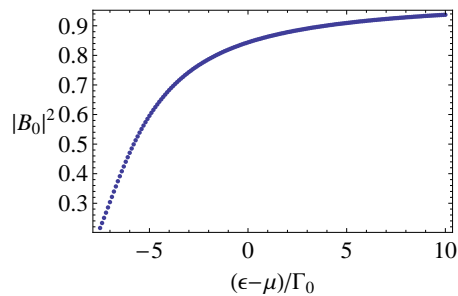}\\
\caption{Holes density $|\mathcal{B}_0|^2$ as a function of $(\epsilon-\mu)/\Gamma_0$ computed setting the remaining parameters as in Fig.\ref{fig:fig2}. } \label{fig:fig3}
\end{figure}

\begin{figure}
%\vspace{3.0cm}
\centering
\includegraphics[scale=0.7]{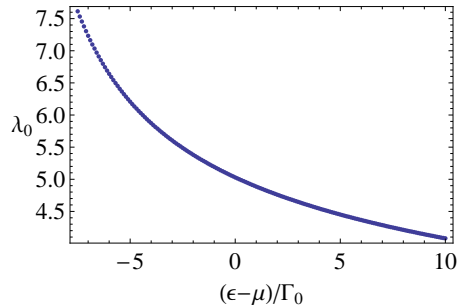}\\
\caption{$\lambda_0$ as a function of $(\epsilon-\mu)/\Gamma_0$ computed setting the remaining parameters as in Fig.\ref{fig:fig2}. } \label{fig:fig4}
\end{figure}

In the following analysis we measure the energy in unit of $\Gamma_0=10 \ \mu eV $, while the current is normalized to the quantity $i_0=-e\Gamma_0/\hbar$ ($i_0 \approx 2.56$ nA). The zero temperature limit is assumed.
In Fig.\ref{fig:fig2} we show the
current pumped in the right lead $I_R$ as a function of the position of the bare level of the dot $\epsilon$ and by setting the remaining parameters as
follows: $\Gamma_{\sigma;0}^{\alpha=L/R}=1.5$, $\Gamma_{\sigma;\omega}^{\alpha=L/R}=0.05$, $\omega=0.05$, $\varphi_L=0$, $\varphi_R=\pi/4$.
Since we are in the spin-insensitive case each spin channel contributes the same quantity to the charge current $I_R$
 ($I_R=I_R^{\sigma}$). The lower curve in Fig.\ref{fig:fig2} represents the contribution to the current induced
 by the modulation of the renormalization of the energy level on the QD (i.e. $\lambda(t)$), the upper curve is
 the current pumped by the modulation of the tunnel barriers, while the middle dashed curve represents the total current which is sum of the previous terms. When $\epsilon \gg \mu$ the current induced by $\lambda(t)$ is a vanishing quantity and the total current is dominated by the modulation of the tunnel amplitudes; on the other hand, when the energy level of the QD $\epsilon$ is placed below the Fermi energy $\mu$, the main contribution to the current is given by the
  modulation of  $\lambda(t)$. From the analysis above we observe a competition between the two different pumping mechanisms which induces a change
  of sign of the total current and presents maximum values in the pA range.
%===================================================================fig2
%=======================================================================
The behavior of the current shown in Fig.\ref{fig:fig2} can be understood by analyzing the slave boson parameters as a function of $(\epsilon-\mu)/\Gamma_0$. Indeed in Fig.\ref{fig:fig3} we plot the $|\mathcal{B}_0|^2$ \textit{vs} $(\epsilon-\mu)/\Gamma_0$ curve fixing the other parameters as in Fig.\ref{fig:fig2}. As expected the electron density on the QD, roughly proportional to $1-|\mathcal{B}_0|^2$, becomes a vanishing quantity when the energy level $\epsilon$ is well above the Fermi energy. In this situation a weak renormalization of the tunneling amplitudes is observed, while in presence of a non-vanishing electron density on the QD the tunneling processes are strongly renormalized. A similar effect is evident in Fig.\ref{fig:fig4} where we present $\lambda_0$ as a function of $(\epsilon-\mu)/\Gamma_0$.
As can be seen the effect of the constraints becomes more pronounced at decreasing $(\epsilon-\mu)$, i.e. when approaching the single occupied state of the dot.
%===================================================================fig3
%=======================================================================

%===================================================================fig4
\begin{figure}[h]
%\vspace{3.0cm}
\centering
\includegraphics[scale=0.7]{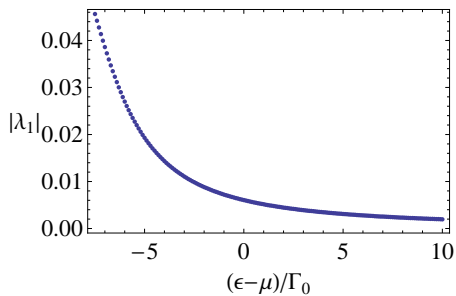}\\
\caption{$|\lambda_1|$ as a function of $(\epsilon-\mu)/\Gamma_0$ computed setting the remaining parameters as in Fig.\ref{fig:fig2}. } \label{fig:fig5}
\end{figure}

\begin{figure}
%\vspace{3.0cm}
\centering
\includegraphics[scale=0.7]{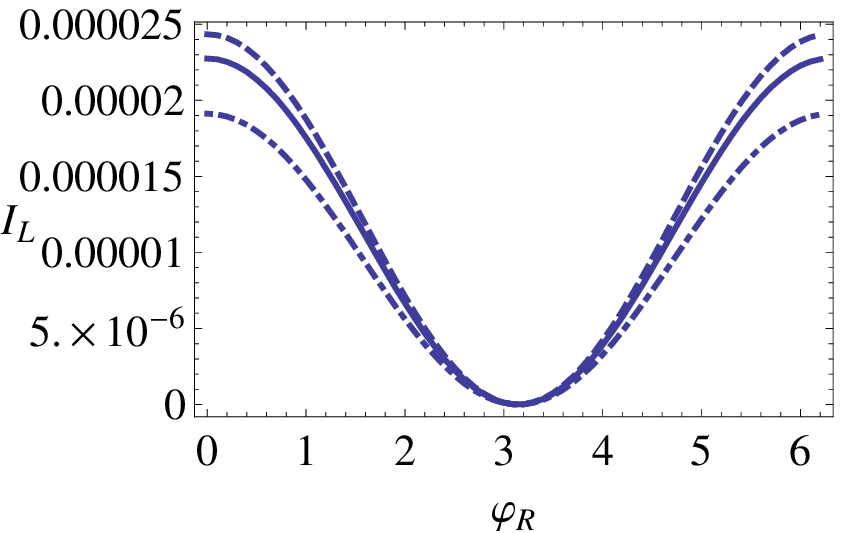}\\
\includegraphics[scale=0.7]{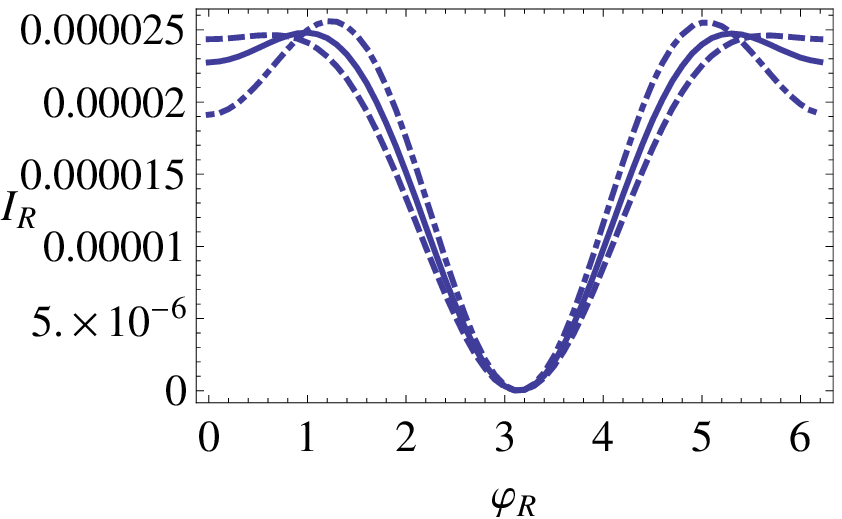}
\caption{Charge current $I_L$ (upper panel) and $I_R$ (lower panel) as a function of the pumping phase $\varphi_R$ computed fixing the remaining parameters as follows: $\Gamma_{\sigma;0}^{\alpha=L/R}=1.5$, $\Gamma_{\sigma;\omega}^{\alpha=L/R}=0.05$, $\omega=0.05$, $\varphi_L=0$. For both the upper and lower panel the dashed curve is computed taking $(\epsilon-\mu)/\Gamma_0=1$, the dashed-dotted curve for $(\epsilon-\mu)/\Gamma_0=-1$, while the full-line is adopted for $(\epsilon-\mu)/\Gamma_0=0$.} \label{fig:fig6ab}
\end{figure}

%=======================================================================
Another effect of the renormalization of $\lambda_0$ is shown in Fig.\ref{fig:fig5} where the absolute value of the oscillation amplitude of $\lambda(t)$ is plotted as a function of the QD bare level. As shown in that figure, below the Fermi energy (i.e. close to $\epsilon-\mu \approx -7 \Gamma_0$) $|\lambda_1|$ becomes comparable to the oscillation amplitudes of the tunneling rates $\Gamma_\sigma^\alpha$
(which approach the value $\approx 0.05 \Gamma_0$) and thus the pump starts to feel the effect of the modulation driven by the internal dynamics of the system. Such effect is evident if we compare Fig.\ref{fig:fig2} and Fig.\ref{fig:fig5}. Indeed Fig.\ref{fig:fig2} shows that the $\lambda$-induced current (lower curve) is enhanced  when $|\lambda_1|$ (see Fig.\ref{fig:fig5}) is close to its maximum value within the parameters range considered.
%===================================================================fig5
%=======================================================================
 In Fig. \ref{fig:fig6ab} we report the dependence of the charge current on the pumping phase $\varphi_R$. It exhibits a
 complicate behavior due to the presence of phase differences, namely $\phi_b^{0,\pm}$ and $\phi_{\lambda}^{\pm}$, induced by the slave boson field and by the modulation of $\lambda(t)$. Due to
 the presence of these phases, the external phase difference $\varphi_R-\varphi_L$ is not the only relevant phase, differently from the noninteracting
 case. Indeed, in the infinite-$U$ limit terms of the form $\sin(\varphi_{\alpha}\pm (\phi_b^{\pm}-\phi_b^{0}))$ appear in the charge current.
 The interaction-induced terms are in general of the same order of magnitude of the ones usually obtained in the free-electron quantum pumping and thus
 strongly affect the usual $\sin(\varphi)$-behavior. Although the charge current contains terms having the usual
 form\cite{wang&wang_cpr} $C+A\sin(\varphi_R-\varphi_L)+B\cos(\varphi_R-\varphi_L)$, its general aspect
 for the parameters range considered here is dominated by a cosine-behavior typical of a rectification contribution, i.e.
 of the form $A+B\cos(\varphi_R)$. The above result is very similar to the one found in Ref.[\onlinecite{burkard-prl}] where the nonadiabatic noninteracting
 case has been considered.
  In the almost adiabatic case treated here, the external ac signals drive the system under non-equilibrium condition inducing an oscillation of the internal
  fields $\mathcal{B}(t)$ and $\lambda(t)$. The latter situation reproduces the external forcing pattern similar to the one proposed in
  Ref.[\onlinecite{burkard-prl}].
%===================================================================fig6ab
%=======================================================================
 The $I_{L/R}$ vs $\varphi_R$ curves present a cosine-dependence, i.e. a rectification contribution, while the precise shape of the curves depends on the energy $\epsilon$ of the QD (see the dashed, dot-dashed and full line in Fig.\ref{fig:fig6ab}).
To render the analysis more quantitative, while maintaining the structure of the equations simple, we explicitly report the expression of the current pumped
in the lead $\alpha$ omitting the current term related to the modulation of $\lambda$:
\begin{eqnarray}
&& \bar{I}^{\alpha}_{\sigma}=-(e/\hbar)\sum_{\eta=\pm}\Bigl \{\sum_{\alpha'}[Re\{\mathcal{R}^{\sigma\eta}_2\}(r_{\alpha}r_{\alpha'}+\iota_{\alpha}\iota_{\alpha'})]\\\nonumber
&& +\eta Im\{\mathcal{R}^{\sigma\eta}_2\}(r_{\alpha}\iota_{\bar{\alpha}}-r_{\bar{\alpha}}\iota_{\alpha})\Bigl \}+(...),
\end{eqnarray}
where $\bar{\alpha}=L \delta_{\alpha R}+R\delta_{\alpha L}$, while the spin-dependent coefficients $r_{\alpha}$ and $\iota_{\alpha}$ are defined as follows:
\begin{eqnarray}
&& r_{\alpha}=\sin(\varphi_{\alpha})\frac{\Gamma_{\sigma;\omega}^{\alpha}|\mathcal{B}_0|^2}{2}+\Gamma_{\sigma;0}^{\alpha}|\mathcal{B}_0||\mathcal{B}_1|\sum_{\eta=\pm}\cos(\phi_b^0-\phi_b^{\eta})\\\nonumber
&& \iota_{\alpha}=\cos(\varphi_{\alpha})\frac{\Gamma_{\sigma;\omega}^{\alpha}|\mathcal{B}_0|^2}{2}+\Gamma_{\sigma;0}^{\alpha}|\mathcal{B}_0||\mathcal{B}_1|\sum_{\eta=\pm}\eta \sin(\phi_b^{\eta}-\phi_b^{0}).
\end{eqnarray}
 As explicitly shown, the charge current contains all the phase differences (i.e. internal or external phase differences) inside the scattering region.  Thus in the strongly interacting case the presence of new phase differences is the main modification compared to the free-fermion
 theory.

\subsection{Spin-selective quantum pumping}
\begin{figure}
\centering
\includegraphics[scale=0.4]{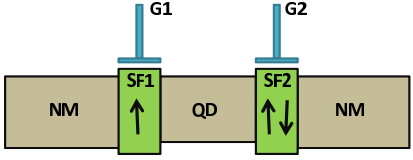}\\
\caption{The spin-sensitive quantum pump described in the main text obtained by connecting via spin-filter barriers (SF) two non-magnetic leads with an interacting quantum dot (QD). An ac modulation of the barrier strength is allowed via top gates G1/G2.} \label{fig:fig7}
\end{figure}

\begin{figure}[h]
\centering
\includegraphics[scale=0.7]{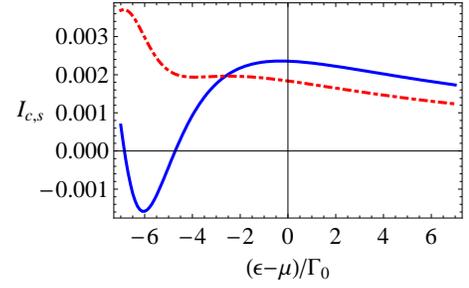}\\
\caption{Charge current $I_c$ (full line) and spin current $I_s$ (dashed-dotted line) pumped in the right lead as a function of $(\epsilon-\mu)/\Gamma_0$ by setting the remaining parameters as follows:
$\Gamma_{0}^{\alpha=L}=1.5$, $\Gamma_{0}^{\alpha=R}=1.3$, $\Gamma_{\omega}^{\alpha=L/R}=0.3$, $\omega=0.05$, $\varphi_L=0$, $\varphi_R=\pi/4$, $p_L=p_R=0.4$.} \label{fig:fig8}
\end{figure}

\begin{figure}
\centering
\includegraphics[scale=0.9]{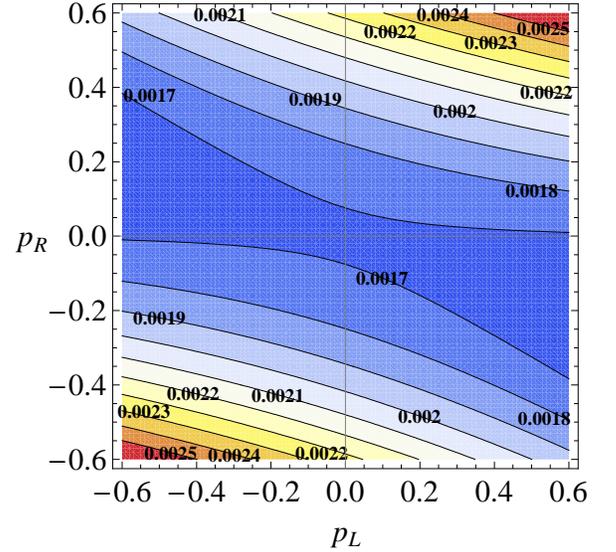}\\
\caption{Contourplot of Charge current $I_c$ as a function of the polarizations $p_{L/R}$ by setting the remaining parameters as follows:
$\Gamma_{0}^{\alpha=L}=1.5$, $\Gamma_{0}^{\alpha=R}=1.5$, $\Gamma_{\omega}^{\alpha=L/R}=0.3$, $\omega=0.05$, $\varphi_L=0$, $\varphi_R=\pi/4$, $(\epsilon-\mu)/\Gamma_0=0$.} \label{fig:fig9}
\end{figure}

\begin{figure}
\centering
\includegraphics[scale=0.9]{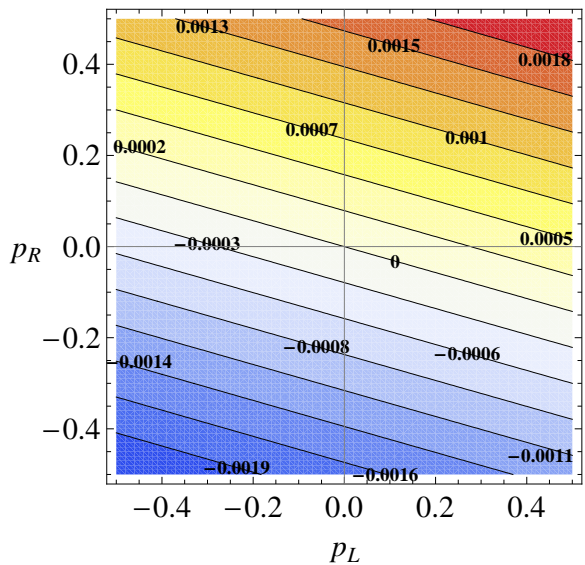}\\
\caption{Contourplot of spin current $I_s$ as a function of the polarizations $p_{L/R}$ by setting the remaining parameters as follows:
$\Gamma_{0}^{\alpha=L}=1.5$, $\Gamma_{0}^{\alpha=R}=1.5$, $\Gamma_{\omega}^{\alpha=L/R}=0.3$, $\omega=0.05$, $\varphi_L=0$, $\varphi_R=\pi/4$, $(\epsilon-\mu)/\Gamma_0=0$. Notice that the relation $I_s\approx A p_L+B p_R$ with $A \approx 0.94 \times 10^{-3}$ and $B \approx 3.3 \times 10^{-3}$ is obeyed.} \label{fig:fig10}
\end{figure}

In this section we study the quantum pumping through the system shown in Fig.\ref{fig:fig7}. It consists of a strongly interacting QD connected to two non-magnetic leads via thin magnetic tunnel barriers. The magnetic barriers work as a spin-filter (SF) whose effect is controlled by the polarization $p_{\alpha} \in [0,1]$ of the $\alpha$-th tunnel barrier. Thus, the spin and charge currents generated by the quantum pumping procedure are strongly affected by relative orientation (parallel or anti-parallel) of the SF barriers. Similar to Refs.[\onlinecite{barnas, souza}], we take the tunneling amplitudes of the following form: $\Gamma_{\sigma}^{\alpha}(t)=\Gamma^{\alpha}(t)(1+p_{\alpha}\sigma)$, where the polarization $p_{\alpha}$ is assumed to be a time-independent quantity, while the tunneling amplitude $\Gamma^{\alpha}(t)$ is taken:
\begin{equation}
\Gamma^{\alpha}(t)=\Gamma_{0}^{\alpha}+\Bigl[\frac{\Gamma_{\omega}^{\alpha}\exp(i\varphi_{\alpha})}{2 i}\exp(i \omega t)+c.c.\Bigl].
\end{equation}
%===================================================================fig7
%=======================================================================
In the presence of the ac modulations of the gates G1/G2 and thanks to the spin-selective properties of the SF,
a spin current $I_s=\sum_{\sigma}\sigma I_{\sigma}$ is generated other than a charge current $I_c=\sum_{\sigma}I_{\sigma}$.
The proposed mechanism which produces a spin polarized current by only using the filtering properties of the magnetic barriers
and avoiding the injection of spin polarized electrons by ferromagnetic leads,
is very similar to the one proposed in Ref.[\onlinecite{miao-prl}] within the dc case and is conceptually new.
In Fig.\ref{fig:fig8} we study the charge (full line) and spin (dashed-dotted line) currents as
 a function of the bare  energy level $\epsilon$ of the QD setting the remaining parameters as in the figure caption. For non-vanishing  polarizations (i.e. $p_L=p_R=0.4$) of the SF barriers a sizable  spin current is observed. Furthermore when $\epsilon \approx \mu-5 \Gamma_0$ the charge current takes negligible values and the pump works as a pure spin current injector.
 A relevant observation is that the pure spin current obtained here is not related to the Zeeman energy of the QD (not present here)
 but is originated by the spin filtering properties of the magnetic barriers connecting the QD to the external unpolarized leads.
%===================================================================fig8
%=======================================================================
Since the only relevant parameters are the polarizations $p_{L/R}$ induced by the SF barriers, in Fig.\ref{fig:fig9} we study the charge current as a function of the relative orientations of $p_L$ and $p_R$ by setting the remaining parameters as follows: $\Gamma_{0}^{\alpha=L}=1.5$, $\Gamma_{0}^{\alpha=R}=1.5$, $\Gamma_{\omega}^{\alpha=L/R}=0.3$, $\omega=0.05$, $\varphi_L=0$, $\varphi_R=\pi/4$, $(\epsilon-\mu)/\Gamma_0=0$.
The analysis of Fig.\ref{fig:fig9} shows the charge current vs the polarization of the SF barriers. It
presents high values of the current within the parallel configuration, while a suppression of the current is observed for the anti-parallel case. This
behavior is explained by observing that for parallel polarizations of the SF barriers only one spin channel is perturbed by
the scattering off the magnetic barrier, while for the antiparallel case both the spin channel are perturbed leading to a less intense current.
%===================================================================fig9
%=======================================================================
The analysis of the spin current $I_s$ performed in Fig.\ref{fig:fig10} shows that the relation $I_s\approx A p_L+B p_R$ with $A \approx 0.94 \times 10^{-3}$ and $B \approx 3.3 \times 10^{-3}$ is obeyed for the considered parameters. The latter observation implies that an intense spin current is obtained within the parallel configuration (positive or negative, depending of the specific sign of the polarizations $p_{\alpha}$), while for the anti-parallel case the spin currents takes less intense values. Very interestingly the spin current can become negligible also for finite values of $p_{L/R}$. Indeed, the above analysis of $I_s$ shows that when the relation $p_L \approx -3.5 p_R$ is fulfilled the spin current goes to zero.
 In particular when the parameters are varied the spin current presents negligible values when $p_L=-\kappa p_R$, where  we have checked that
 the coefficient $\kappa$ depends on the position of energy level $\epsilon$ of the QD.
 Varying it one can rotate the line $p_L=-\kappa p_R$ within the $(p_L,\ p_R)$ plane.
%===================================================================fig10
%=======================================================================
This property shows an all-electrical control of the spin current, potentially useful in spintronics devices.

\section{Conclusions}
\label{sec:conclusions}
We studied the charge and spin current pumped in an interacting quantum dot connected to two external leads via tunnel barriers
whose transparencies are periodically modulated in time via the top gates G1/G2.
The current has been calculated by a non-equilibrium Green's function (GF) method based on a time dependent slave boson approach.
In particular, we derived an equation of motion
of the slave boson field $\mathcal{B}(t)$ in terms of an infinite series of constraints, generalizing the adiabatic case.
By using a finite set of constraints (single photon approximation)
the expressions of the relevant Green's functions and of the current have been derived.
We showed the numerical results of the pumped current for two different situations: the spin insensitive and the spin dependent case. In the latter case the spin selective properties of the system are determined by the presence of two magnetic tunnel barriers.
In the spin insensitive case the system pumps a charge current that contains beyond a term related to the modulation of the out-of-phase external parameters an additional term due to the internal dynamics of the Lagrange multiplier and the slave boson field. This term is identified by us as a parasitic pumping current. Its presence is due to the phase differences inside the scattering region and is a consequence of strong correlations effects. In fact, the charge current, which for a free-fermions model depends only on the phase difference between the two external parameters, $\varphi_R-\varphi_L$, in the strongly interacting limit depends separately on $\varphi_{L/R}$. The additional pumping contributions are originated by the phase difference of the external parameter phase (for instance $\varphi_L$) and the phase of an internal degree of freedom (i.e. the slave boson field or the Lagrange multiplier). This mechanism not only works in the interacting quantum dot case but can operate in the presence of any internal dynamics, as pointed out in Ref.[\onlinecite{deformable-qd}, \onlinecite{memory-pumping}].\\
Concerning the spin sensitive case the current preserves the above characteristics, while the appearance of a non vanishing spin current is shown.  The spin current is of the form $I_s \approx A p_L+B p_R$, where the parameters $A$, $B$ depend on the electronic configuration of the QD and $p_L,p_R$ are the polarizations of the magnetic tunnel barriers. By considering specific values of the polarization of the barriers, a change of sign of the spin current can be obtained by tuning the energy level $\epsilon$ of the QD. This produces a rotation of the line $I_s(p_L,p_R)=0$ in the $(p_L,p_R)$ plane. Furthermore, acting on the energy level of the QD the system can be driven toward a working point in which the pump works as a pure spin current injector. These two mechanisms shows the possibility of all-electrical control of the spin current pumped through the system relevant in spintronics devices.

%\section*{ACKNOWLEDGMENTS}

\appendix{}
\section{Definition of $\Phi^{(n;\eta)}(E)$}\label{app:fi-function}
The definition of the function $\Phi^{(n;\eta)}(E)$ in terms of the GFs of the quantum dot region takes the following form:
\begin{eqnarray}
&&\Phi^{(n;\eta)}(E)=\sum_{\sigma}\int\frac{dE_2}{(2\pi)^2}W^{(n)}_{\sigma}\\\nonumber
&\times&[f(E_2-n\omega)G_{\sigma;\eta}^{r}(E+E_2-n\omega)\\\nonumber
&+&\frac{1}{2}G^{<}_{\sigma;\eta}(E+E_2-n\omega)],
\end{eqnarray}
where the function $W^{(n)}_{\sigma}$ is defined as follows:
\begin{eqnarray}
&& W^{(n=0)}_{\sigma}=\Gamma_{\sigma;0}\mathcal{B}_0\\\nonumber
&& W^{(n=\pm1)}_{\sigma}=\Gamma_{\sigma;0}\mathcal{B}_{\pm 1}+\Gamma_{\sigma;\pm 1}\mathcal{B}_0,
\end{eqnarray}
with $\Gamma_{\sigma;n}=\sum_{\alpha}\Gamma_{\sigma;n}^{\alpha}$.

%=======================================================================================bib
\bibliographystyle{prsty}

\begin{thebibliography}{99}
\bibitem{nota1} The infinite $U$-limit is appropriate for quantum dots characterized by $U \sim 1$ meV and transition rate $\Gamma_0 \sim 1-10\mu $eV.
\bibitem{nota1b} The notation $\bar{\sigma}=-\sigma$, being $\sigma=\pm1$ the up/down ($\uparrow \equiv 1 $, $\downarrow \equiv -1$) components of the spin projection, is adopted in the main text.
\end{thebibliography}

\end{document}